\documentclass[3p,times,twocolumn]{elsarticle}

\usepackage{ecrc}
\usepackage{graphicx}
\usepackage{epsfig}
\usepackage{subfigure}
\usepackage{amsmath}

\volume{00}

\firstpage{1}

\journalname{Nuclear Physics B Proceedings Supplement}

\runauth{}

\jid{nuphbp}

\jnltitlelogo{Nuclear Physics B Proceedings Supplement}

\usepackage{amssymb}

\usepackage[figuresright]{rotating}

\begin{document}

\begin{frontmatter}

\title{Status and Perspectives for $\bar PANDA$ at FAIR}
\author{Elisabetta Prencipe} 
\author{on behalf of the $\bar PANDA$ Collaboration}
\ead{e.prencipe@fz-juelich.de}
\address{Forschungszentrum J\"ulich, Leo Brandt Strasse, 52428 J\"ulich, Germany}

\dochead{}


\begin{abstract}
The Facility for Antiproton and Ion Research (FAIR) is an international accelerator facility which will use antiprotons and ions to perform research in the fields of nuclear, hadron and particle physics, atomic and anti-matter physics, high density plasma physics and applications in condensed matter physics, biology and the bio-medical sciences. It is located at Darmstadt (Germany) and it is under construction. Among all projects in development at FAIR in this moment, this report focuses on the $\bar PANDA$ experiment (antiProton ANnihilation at DArmstadt). Some topics from the Charm and Charmonium physics program of the $\bar PANDA$ experiment will be highlighted, where $\bar PANDA$ is expected to provide first measurements and original contributions, such as the measurement of the width of very narrow states and the measurements of high spin particles, nowaday undetected. The technique to measure the width of these very narrow states will be presented, and a general overview of the machine is provided.

\end{abstract}

\begin{keyword}
$D_S$ \sep Charm \sep Spectroscopy \sep confinement \sep $\bar PANDA$

\end{keyword}

\end{frontmatter}

\section{Introduction}
\label{Introduction}
The Standard Model of particle physics is well defined and efficient in describing fundamental interactions.  However several questions still remain open. For example, the theory describing strong interactions, the Quantum Chromodynamcs (QCD), is still affected  by some unsolved fundamental questions, arising in the low energy domain, such as the understanding of confinement and the origin of hadron masses. 
As a non-Abelian theory, QCD allows the self-interaction of the strong force carriers, e.g. the gluons. In the low enery regime their interactions can only be described exploiting non-perturbative methods. 
The answer to these questions is a challenge that requires a new generation machine and experiments with higher resolution and better precision, compared to the past.\\
The future experiment $\bar PANDA$ will be located at the HESR at FAIR\cite{fair} (High Energy Storage Ring at the Facility for Antiproton and Ion Research), in Germany.

In this report we will put emphasis on the description of the $\bar PANDA$ experiment, in particular on the detector design and the physics program, to motivate the big effort in terms of hardware and software that an international collaboration of 18 countries and more than 500 people are presently going through.

\section{The physics case}
\label{Physics}

The program of the $\bar PANDA$ experiment is wide and ambitious, and covers several areas of interest in nuclear and particle physics\cite{pandabook}.
 
We plan to study with accuracy the mechanism responsible for phenomena like the quark confinement, through the investigation of:
\begin{itemize}
\item Hadron spectroscopy:\\
$-$ search for gluonic excitations;\\
$-$ charmonium spectroscopy;\\
$-$ D meson spectroscopy;\\
$-$ baryon spectroscopy;\\
$-$ QCD dynamics. 
\item Nucleon structure:\\
$-$ parton distribution;\\
$-$ time-like form factors of the proton;\\
$-$ transition distribution amplitudes.
\item Hadrons in matter.
\item Hypernuclei.
\end{itemize}

$\bar PANDA$ is a fix-target experiment, where a beam of antiprotons will collide against a thick target, e.g. 4$\cdot$10$^{15}$ cm$^{-2}$, with a beam life time $>$ 30 minutes. The choice of an antiproton beam is strongly supported from the fact that the $\bar p p$ interactions are gluon rich processes. All quantum numbers will be directly accessible in the annihilation. Several  advantages are available in this respect:
\begin{itemize} 
\item very good mass resolution, which  depends basically on the beam resolution and not on that of  the detectors. A 100 keV pitch mass scan will be possible in $\bar PANDA$, which is 20 times better than attained at B factories and more than 2 times better than at the Fermilab experiment E760;
\item  direct formation of high spin states, forbidden at B factories;
\item  direct production with very high rate/day.
\end{itemize}
\section{The detector}
\label{detector}
\begin{figure*}[ht] 
\begin{center}
\mbox{
{\scalebox{0.5}{\includegraphics{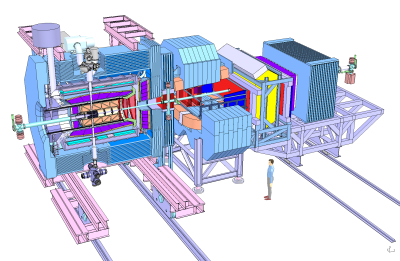}}}}
\mbox{
{\scalebox{0.25}{\includegraphics{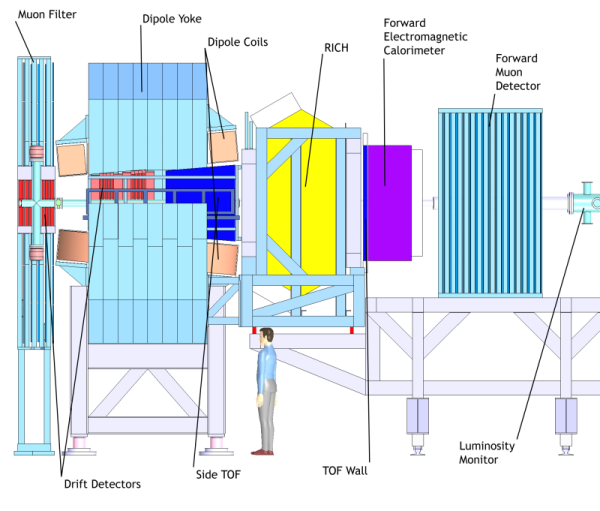}}}\quad
{\scalebox{0.53}{\includegraphics{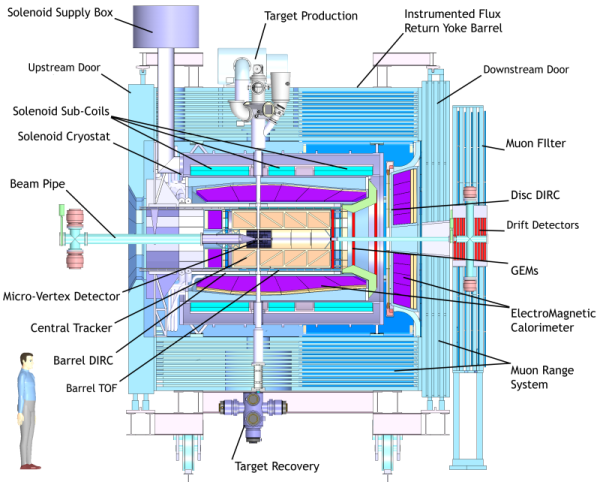}}} 
}
\caption{\label{Fig1-panda} General view of the $\bar PANDA$ detector (top). Detailed view of the central spectrometer (down-left) and forward spectrometer (down-right) of $\bar PANDA$.}
\end{center}
\end{figure*}
\begin{table*}
\caption{Operation modes of $\bar PANDA$ with HESR.}
\label{table1-panda}
\begin{center}
\vskip -0.2cm
\begin{tabular}{lrcccl}
\hline \hline 
        &High resolution mode  &High luminosity mode \cr \hline
cooling & e$^{-}$,  1.5 $\leq$ p $\leq$ 8.9 GeV/c&stochasting, p $\geq$ 3.8 GeV/c \cr
N. of anti-protons stored                & 10$^{10}$                &10$^{11}$   \cr
Luminosity [cm$^{-2}$ s$^{-1}$]  & $\leq$ 2$\cdot$10$^{31}$ &$\leq$ 2$\cdot$10$^{32}$  \cr
$\Delta$p/p                    & 4$\cdot$ 10$^{-5}$       &2$\cdot$ 10$^{-4}$\cr \hline
\end{tabular}
\end{center}
\end{table*}
 The fixed-target experiment $\bar PANDA$ is composed by two main parts: the central and the forward spectrometer, as shown in Fig.~\ref{Fig1-panda}, inserted in a  homogeneous solenoid magnetic field ($B$ = 2T) and a dipole field ($B$ = 2T$\cdot$m), respectively. Two options are still under investigation for the target: cluster-jet or pellet target. $\bar PANDA$ will span a wide momentum range, from 1.5 up to 15 GeV/c. Focalized through stochastic and electron cooling, the antiproton beam will have excellent momentum resolution. Two operation modes are provided: high resolution mode  and  high luminosity mode, as reported in Table ~\ref{table1-panda}.

$\bar PANDA$ is a 4$\pi$-coverage machine.
The $\bar PANDA$ innermost detector is the Micro-Vertex-Detector (MVD), a sophisticated silicon pixel and silicon strip array, which will provide a good vertex reconstruction, essential to reduce the high level of background, and to reconstruct lower momentum particles. A vertex space resolution of 50 $\mu m$ in x,y, and 100 $\mu m$ in z is expected. The GEM and the Straw-Tube-Tracker (STT) will allow to track charged particles ($\Delta$p$\rm _T$/p$\rm _T$ = 1.2$\%$ together with the MVD). A Cherenkov detector (DIRC) is planned to discriminate with excellent efficiency K/$\pi$. A high performant calorimeter will be equipped with 17200 $PbW0_4$ crystals, operating at the temperature of -25$^0$ $C$; it will be provided of 2 ADPs, so the performances are definitively better than that of the CMS experiment. It will provide an excellent separation between pions and electrons, and excellent photon reconstruction.  A forward system is needed to track particles which will be emitted ahead due to the high boost of proton-antiproton in the center of mass.  The muon detector, together with a luminosity monitor, are the outmost detectors of $\bar PANDA$.  

Background and interesting signal events in $\bar PANDA$ will have the same signature; therefore no hardware trigger is able to discriminate $a$ $priori$  background. Online reconstruction can be exploited (``software'' trigger): considering that detector acceptance only, the ratio between  signal and background (S/B) is expected to be 10$^{-6}$, with 20 MHz average interaction rate. This number makes  many searches of the $\bar PANDA$ physics program challenging. $Ad$ $hoc$ techniques to reject the high background level are necessary, depending on the physics channel.

In this report, some simulations performed within the $PandaRoot$\cite{pandaroot1, pandaroot2}\footnote{$PandaRoot$ is the official $\bar PANDA$ framework in development inside the project FairRoot@GSI.} framework will be reported, to show the healthy status of our software project development and to point out which are the potentialities and the original contributions expected from $\bar PANDA$ in the field of Charm and Charmonium spectroscopy. 

\section{Challenges in Charm physics with $\bar PANDA$}
\label{charm}

\begin{figure}[h] 
\begin{center}
\scalebox{0.3}{\includegraphics{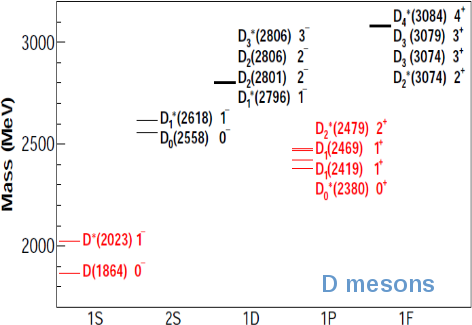}}
\scalebox{0.3}{\includegraphics{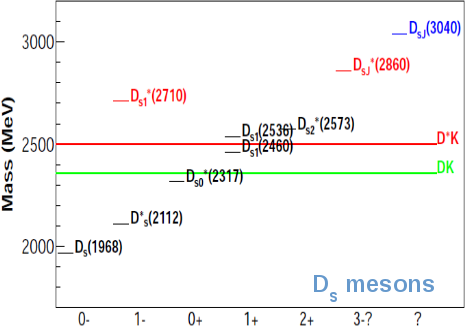}}
\caption{\label{Fig3-panda} Mass spectrum versus the particle spin J$\rm ^{PC}$ for D mesons (top) and  D$_s$ mesons (down). The horizontal long lines indicate the DK and D$^*$K threshold; the short horizontal lines indicate the predicted states as in Ref.~\cite{dipierro1, dipierro2}. The mass spectra here reported are built using recent experimental results, presented at the conference CHARM2013 and available in Ref.~\cite{charm2013}.}
\end{center}
\end{figure}
The study of D mesons is important both for strong and weak interactions. Gluonic excitations and hadrons composed of strange and charm quarks can be abundantly produced in $\bar p p$ interactions and their features will be accessible with unprecedented accuracy, thereby allowing high precision tests of the strong interaction theory in the intermediate energy regime. On the other hand, the search for CP violation in the D sector recently has gained more attention, as a new field of investigation.
 
Understanding the $cs$-spectrum (see Fig.~\ref{Fig3-panda}) is not easy: 11 years after the discovery of the charged state called $D_{s0}^*(2317)$\cite{Dsbabar}, its mass is known with high precision, but for its width only an upper limit exists. The observation of the  $D_{s0}^*(2317)$ represents a break-point, because the existence of combined $c$,$s$ quark systems is theoretically predicted\cite{dipierro1, dipierro2}; but some experimental observations questioned the potential models, which fairly agree with the observed D meson spectrum, e.g. mesons composed of the light quark $u$ or $d$ and the heavy quark $c$. Potential models agree also with the observation of several $D_s$ states, up to the discovery of the $D_{s0}^*$(2317).  But they cannot explain why the $D_{s0}^*(2317)$ mass was observed more than 100 MeV/c$^2$ below the predictions. This is an even more complicated issue, as $D_s$ mesons are composed by the quarks $s$ and $c$ (see Fig.~\ref{Fig3-panda}). The same is valid for the observation of the so called $D_{s1}(2460)$\cite{Ds2460}: its mass was found below the theoretical expectations as well. $D_{s0}^*(2317)$ and $D_{s1}(2460)$ are both observed below the DK threshold; they are very narrow, and their decays are isospin violating; it is very difficult to predict the cross section of $\bar p p \rightarrow D_s^{+(*)} D_s^{-(*)}$, as we cannot perform perturbative calculations, since they would underestimate the real cross section. The quantum numbers of $D_{s0}^*(2317)$ and $D_{s1}(2460)$ are, in any case, not fixed yet, although we can exclude $J{^P}$=0$^+$ for the $D_{s1}(2460)$ based on experimental observations\cite{widthgg}. In order to answer the important questions related to their interpretation, we need to measure their widths  ($\Gamma$), because they may allow to discriminate among different theoretical models, which provide and explanation for the $D_s$ excited states as  pure $cs$-state ($\Gamma \sim$10 keV)\cite{purecs}, or tetraquark ($\Gamma$ in the range of 10 $-$ 100 keV)\cite{tetraquarks}, or molecular state ($\Gamma \sim$130 keV)\cite{molecules}, or chiral partners of the same heavy-light system built with one heavy and one light quarks ($c$ and $s$, respectively). 

In the past two years the experiment LHCb\cite{charm2013} improved the knowledge of  the $D_s$ spectrum, and confirmed the previous measurements, with the highest world precision; but it cannot provide the measurement of the width of these very narrow states, so the nature of the $D_s$ excited states still remain unclear. With the fine mass scan techniques every 100 keV, $\bar PANDA$  is in a unique position to perform the study of the excitation function of the cross section (see Eq.~\ref{eq1-panda}), and discriminate among the theoretical models. The cross section at given energy $\lambda$ is given by\cite{christoph2}:
\begin{equation}
\label{eq1-panda}
\sigma(\lambda) = \sqrt{m_R \Gamma}  \cdot \lvert \mathcal{ M^{\rm 2}} \rvert  \cdot \frac{1}{\pi} \int_{- \infty}^{\lambda} \frac{\lambda -x}{x^{2} - 1} dx , 
\end{equation}
\begin{equation}
\label{eq2-panda}
\sigma(0) = \sqrt{m_R \Gamma / {\rm 2}} \cdot \lvert \mathcal{ M^{\rm 2}} \rvert
\end{equation}
where $\mathcal{M}$ is the matrix element, $m_R$ is the resonance mass and $\Gamma$ its width, $\lambda$= ($\sqrt s - m_R - m_{Ds}$)/$\Gamma$, $\sqrt s$ = energy in the center of mass for the production e.g. of $D_{s0}^*(2317)$ in the process $\bar p p \rightarrow D_s^- D_{s0}^{*+}$(2317),  that is equal to 4.286 GeV/c$^2$ at the threshold of this process.

\begin{figure*}[ht] 
\begin{center}
\mbox{
{\scalebox{0.259}{\includegraphics{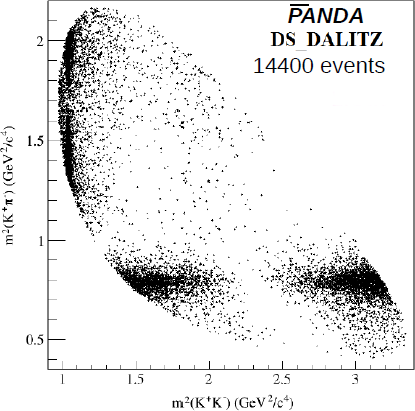}}} \quad
{\scalebox{0.35}{\includegraphics{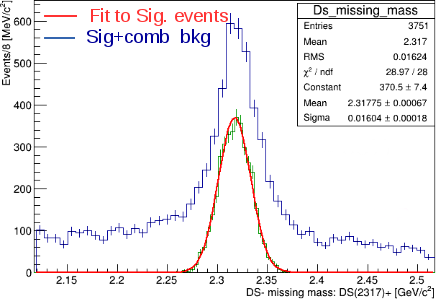}}} \quad
{\scalebox{0.35}{\includegraphics{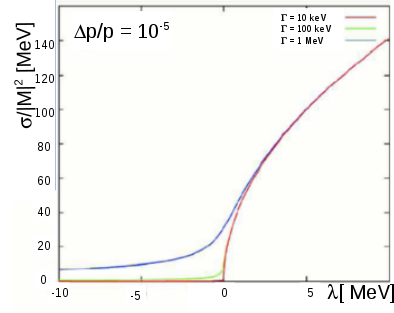}}}
}
\caption{\label{Fig4-panda} Dalitz plot of $\rm m^2_{K^+K^-}$ vs  $\rm m^2_{K^+ \pi^-}$ on simulated $\bar PANDA$ events with the MC generator EvtGen: the realistic CLEO model is used for the reconstruction of the $D_s^- \rightarrow K^+K^- \pi^-$\cite{lucao} (left). MC simulations of $D_{s0}^{*}$(2317)$^+$, reconstructed as  missing mass of the event in the process  $\bar p p \rightarrow D_s^{-} D_{s0}^{*}$(2317)$^+$,  $D_s^- \rightarrow K^+K^- \pi^-$ and $D_{s0}^{*}$(2317)$^+ \rightarrow D_s^+ \pi^0$ (center). Excitation function of the cross secton of the same process as in (b): the curve is reconstructed scanning the $D_{s0}^{*}$(2317)$^+$ every 100 keV, and representing every point of the mass scan in a graph of the cross section as function of the energy difference $\lambda$ = ($\sqrt s - m_{Ds(2317)} - m_{Ds}$)/$\Gamma$ (right).}
\end{center}
\end{figure*}

Studies are planned  in $\bar PANDA$ for the process $\bar p p \rightarrow D_s^- D_s^{*+}$, where $D_s^{*+}$ stands for $D_{s0}^*(2317)$, $D_{s1}(2460)$, and $D_{s1}'(2535)$. There are manifold interests in these decay processes:
\begin{itemize}
\item calculate the cross section of the process $\bar p p \rightarrow D_s^{+(*)} D_s^{-(*)}$ (difficult to predict: expected in the range [1$-$100] nb);
\item study of mixing between $D_s^{(*)}$ states; 
\item measure the width of the $D_{s0}^*(2317)$ and $D_{s1}(2460)$, that will  be first observation. The $D_{s1}'(2535)$ is above the DK threshold, and its width is known with large uncertainty\cite{Ds2535belle, Ds2535babar}; therefore it could be repeated in $\bar PANDA$ with higher precision. It would represent an important cross check of our analysis technique. At threshold  Eq.~\ref{eq1-panda} reduces to the Eq.~\ref{eq2-panda}, so the only observables which we should measure in this case are the mass of the resonant state, and the cross section. 
\end{itemize}
Recent $\bar PANDA$ simulations, shown in Fig.~\ref{Fig4-panda}, are performed using the MC generator EvtGen\cite{evtgen}, within the $PandaRoot$ framework. We use the same model described in Ref.~\cite{Dsbabar}, a Dalitz model based on real data, to simulate a realistic case and estimate the run-time needed in $\bar PANDA$. The high peformance tracking detectors of $\bar PANDA$ allow excellent track reconstruction and high K/$\pi$ separation, together with the DIRC; the designed vertex detector allows high background rejection, by setting tight topological selection cuts around the fitted vertex. We reconstruct the $\rm D_s$ by using a vertex fit with three charged particles, which are identified as K or $\pi$ by means of a likelihood PID (Partile Identification)  method, that makes use of several variables like  energy loss, Cherenkov angle, Zernik momenta. Track finder and fitting procedures in the central spectrometer use the Kalman filter method; wherever the $B$ field is not homogeneous, the Runge Kutta track representation is used. In order to have better mass resolution and higher reconstruction efficiency, the  missing mass of the event is exploited: the reconstruction of the $\rm D_s^-$ is performed, then we extract information on the  $D_{s0}^*(2317)$ by evaluating its four-momentum as the difference between the reconstructed $\rm D_s^-$ four-momentum and the initial state vector. With this technique, we have obtained $\sim$30$\%$ reconstruction efficiency. A full simulation, including electronics and detector material, is being performed. In order to reject naively part of the huge background, originating from very low momentum particles, a preselection cuts on the track momentum  p$_{TRACK}>$100 MeV/c and on the photon momentum p$_{\gamma}>$50 MeV/c are applied in our simulations. A challenge of this analysis is the reconstruction of the many low momentum pions. This is the first time a full simulation including the $D_{s0}^*$(2317) is achieved with $PandaRoot$: the work in still in progress. Our simulations are based on Geant3\cite{geant3}.  With the naive selection cuts just described, the ratio S/B improved from  10$^{-6}$ to 10$^{-2}$. The optimization of the final selection criteria is presently ongoing. 
 The measurement of the width  will be an original and extremely important $\bar PANDA$ contribution to solve the $cs$-spectrum puzzle. In Fig.~\ref{Fig5-panda}c (right) the plot related to Eq.\ref{eq2-panda} is shown, scanning the $D_{s0}^{*}$(2317)$^+$ mass in 100 keV steps around its nominal value: the shape of the curve changes, depending on the input width given to the simulation. The minimum momentum needed to produce the $D_{s0}^*(2317)$ in the process mentioned above is $p$=8.8 MeV/c.\\
In high luminosity mode (see Table~\ref{table1-panda}), with a cross section in the range [1$-$100] nb, and assuming a luminosity of 8.64 pb/day, with a reconstruction efficiency of 30$\%$, we estimate with $\bar PANDA$ a $D_{s0}^*(2317)$ production rate in the order of (3$-$300)$\cdot$10$^3$ per day, to be scaled by the Branching Fraction (BF) of the reconstructed $D_s$. 

\section{Future measurements in the Charmonium sector with $\bar PANDA$}
\begin{figure*}[ht] 
\begin{center}
\mbox{
{\scalebox{0.20}{\includegraphics{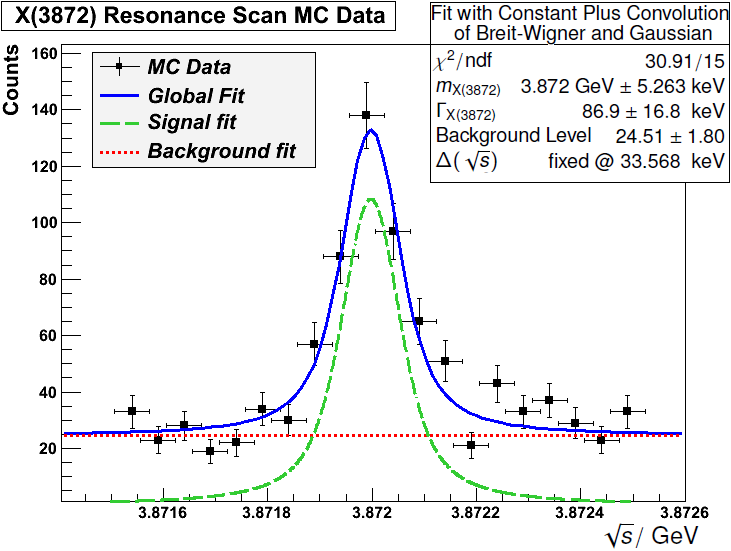}}} \quad 
{\scalebox{0.30}{\includegraphics{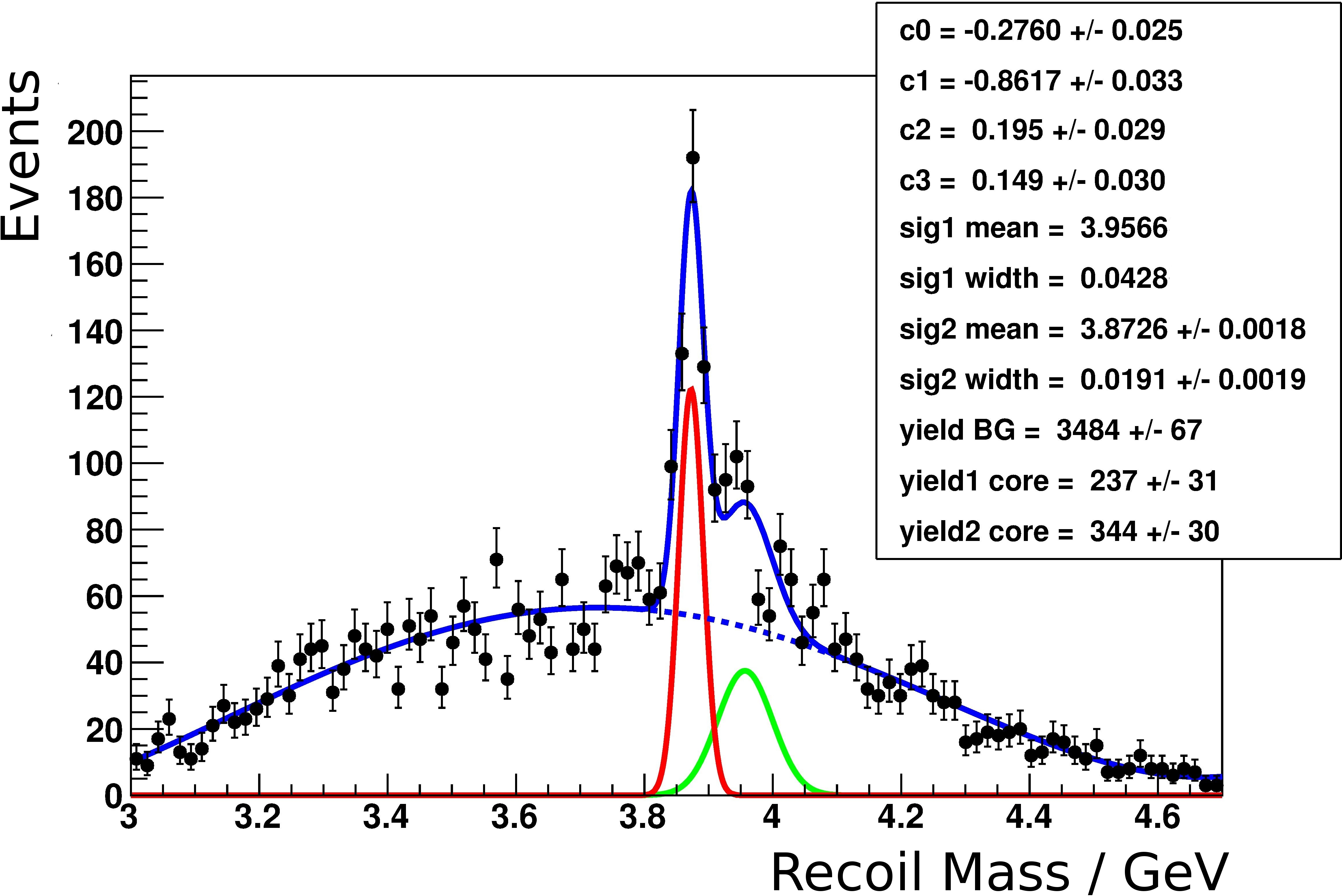}}} \quad   
{\scalebox{0.36}{\includegraphics{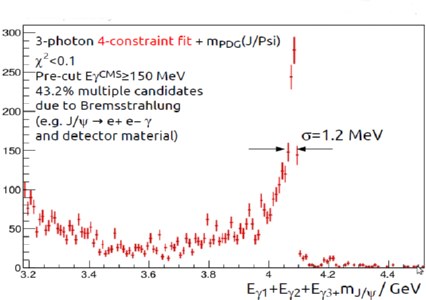}}}        
}
\caption{\label{Fig5-panda} Full $PandaRoot$ simulations  of $X(3872)$ (left), reconstructed in the process $\bar p p \rightarrow X(3872)$, $ X(3872) \rightarrow J/\psi \pi^+ \pi^-$; $h_c'$ (center), reconstructed as recoil of the di-pion mass in the process $\bar p p \rightarrow  \pi^+ \pi^- h_c'$, $h_c' \rightarrow D^0 \bar D^{0*}$ and $D^0 \rightarrow K^- \pi^+$; $^3F^4$ (right), directly formed in the process $\bar p p \rightarrow (^3F^4)$, as radiative $J/\psi$ cascade\cite{Xtogether}.}
\end{center}
\end{figure*}
In 2003 a new era started not only for the charm, but also for the charmonium sector: the discovery of the X(3872)\cite{belleX} and its subsequent confirmation by several experiments (at $e^+e^-$ colliders and $pp$ interaction machines) and in several decay modes. These observations set up the starting point to look after new forms of aggregation of matter: tetraquarks, molecular states, hybrids, etc.\cite{yellow}. These states were predicted by theory even before the discovery of the X(3872), but in the last decade a plenty of theoretical papers provided new interpretations and new models, that have spurred  refreshed interest in spectroscopy. Several new observations and controversial interpretations of the evidence of additional new resonant states suggest that limitations due to statistics and resolution do not allow a unique interpretation, and discriminate among theoretical models. We use to name as generically X, Y, Z  these new experimental findings, which do not fit in the spectra forseen by potential models, as in some cases it has been difficult to assign the quantum numbers. Lots of progresses have been made by LHCb; we know that the X(3872) is an isospin violating and very narrow state; its quantum numbers have been found to be 1$^{++}$\cite{lhcbX}, and no charged partners are found so far. Clearly the interpretation of the X(3872) as charmonium state is unlikely, and alternative models have to be used to understand its nature. As for the case of the $D_{s0}^*$(2317), the measurement of the width plays an important role.

$\bar PANDA$ will scan the X(3872) mass in 100 keV steps. In Fig.~\ref{Fig5-panda} (left) the simulated X(3872) signal is shown, together  with a realistic background; the latter is obtained  using DPM\cite{dpm-panda} MC generator, while the signal events are simulated by EvtGen within the $PandaRoot$ framework: we could reproduce precisely the input values of this simulation in the fit of Fig.~\ref{Fig5-panda}. A detailed description of this work is provided in \cite{martin}. 
\begin{figure}[h] 
\begin{center}
\mbox{
{\scalebox{0.26}{\includegraphics{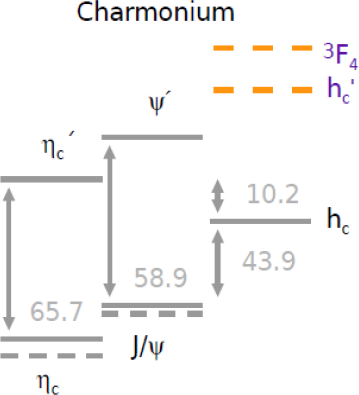}}} \quad 
{\scalebox{0.26}{\includegraphics{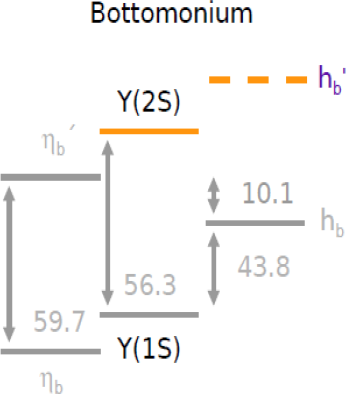}}} \quad
}
\caption{\label{Fig6-panda} Splitting levels of Charmonium (left) and Bottomonium states (right). Similarities are shown in the charm and bottom sector. The dashed lines indicate predicted but undetected resonant states as explained in Ref.~\cite{Xtogether}.}
\end{center}
\end{figure}
The measurement of the X(3872) width is just an example of the original measurements that $\bar PANDA$ will be able to perform in the Charmonium sector. One of the challenges of $\bar PANDA$, for instance, will be to explore and understand the confinement of quarks. A way to do that is to verify and test the behaviour of the potential in the intermediate and high energy regime with heavy hadrons, e.g. those composed particles whose mass is above the $D \bar D$ threshold, since below it the behaviour of the hadrons is generally well predicted from the potential models. A chance to test the confinement is, for example, to search for a $^3F_4$ state\cite{theory3f4}. If this state is found, it could be the proof that the linear term of the potential, in the formula $V(r) = - \frac{4}{3} \frac{\alpha_s}{r} + kr$, is indeed logarithmic, and not linear as it is believed. In fact, we could consider that the charmonium splitting is roughly the same as for bottomonium states (see Fig.~\ref{Fig6-panda}). This can be explained assuming that the potential, at high energy, is logarithmic\cite{Xtogether}. The search for a high spin particle such as $^3F^4$  is also forbidden at B factories (the past BaBar and Belle, or the future Belle II), and it is forbidden at BES III. But $\bar PANDA$ will have easy access to high spin states in formation, with excellent high production rates. In Fig.~\ref{Fig5-panda} preliminary full MC simulations with $PandaRoot$ are performed, to search for two of these states to test the charmonium splitting hypothesis. 

\begin{figure*}[ht] 
\begin{center}
\mbox{
{\scalebox{0.205}{\includegraphics{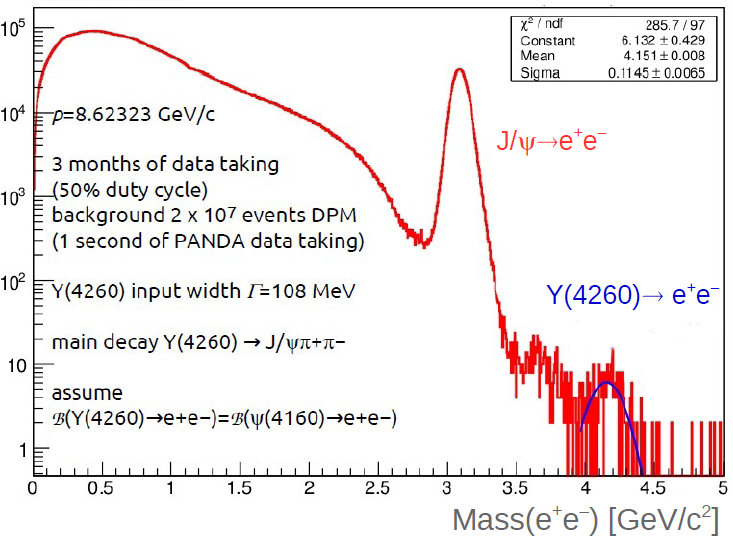}}} \quad 
{\scalebox{0.43}{\includegraphics{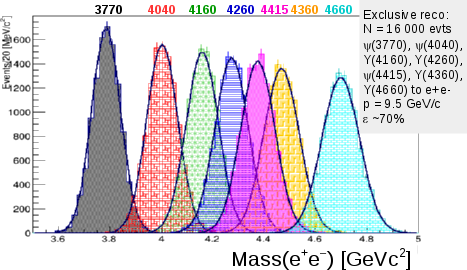}}} \quad
}
\caption{\label{Fig7-panda} Full simulation of vector states above the $\bar D D $ threshold in $\bar PANDA$. Background simulation with DPM, with a clear $J/\psi$ and $Y(4260)$ signal (left); reconstruction of known Y-vector states to $e^+e^-$ with EvtGen, inside the $PandaRoot$ framework, to evaluate the reconstruction efficiency. It is evident that they overlap and inteference effects are expected (right).}
\end{center}
\end{figure*}

A last study illustrated in this report is the simulation of several unpredicted vector resonant states above the  $D \bar D$ threshold, with large width and with established quantum numbers (1$^{--}$), whose nature is still unclear. Fig~\ref{Fig7-panda} shows a $PandaRoot$ full simulations for six of them: the VLL PHOTOS\cite{evtgen} model is used. We know that the BF of the $J/\psi \rightarrow e^+e^-$ is $\sim$6$\%$, and that is the narrow charmonium ground state; but the BF of the Y-family states decaying to $e^+e^-$ is of order of 10$^{-6}$. This could be explained assuming that the radial quantum number $n$ for those Y-states is larger than 1, fact that supports the idea that they are non-conventional charmonia. However, only few of these vector states seem to decay to $e^+e^-$.  For example, no measurement exists of BF($Y(4260) \rightarrow e^+e^-$). This could be due to the statistics limitation at B factories, or because the vector state $Y(4260)$ is not a charmonium state, as in this case it would decay to $e^+e^-$. There are arguments in favor of the $Y(4260)$ as hybrid; but its nature is not fixed, yet. It is worth in any case to look for the decay of all these vector states to $e^+e^-$. Simulations in $\bar PANDA$ are very promising in this sense, as the reconstruction efficiency is evaluated to be $\sim$70$\%$ for each state decaying to $e^+e^-$, and the background simulations by DPM show that background rejection is easy when the center of mass energy of $e^+e^- $ is larger that 4 GeV. We expect roughly 16 000 events/day in the high luminosity mode: this is an incredible high rate compared to what was measured in the past experiments. In addition, we could also study interference effects among all these large vector states, which overlap in a narrow energy range: this study was never performed before.     
\section{Conclusions}
The Standard Model is solid, but it does not give answers to all questions. Several open issues still exist in the Charm and Charmonium physics sector, for example. The $\bar PANDA$ experiment at FAIR is designed to achieve  a mass resolution 20 times better than attained at B factories, essential to perform fine mass scan (every 100 keV). $\bar PANDA$  will have unprecedent PID power, indispensable for the measurement of the proton form factors and other measurements.  \\
In this report we highlighted  the measurement of the width of some Charm and Charmonium states. This task is challenging but the measurement is absolutely needed to clarify and understand the nature of some observed  states, such as the $D_{s0}^*(2317)$, $D_{s1}(2460)$ and X(3872). Understanding the confinement is also a challenge: a new approach could be based on  looking for high spin resonant states to test the potential models.  $\bar PANDA$ offers a unique opportunity to perform these measurements with high precision. \\
TDRs of several detectors have already been approved, and tests with detector prototypes are ongoing. The official $PandaRoot$ simulation framework is at an advanced stage. Important contributions are expected from $\bar PANDA$ when it will start to collect data.

\nocite{*}
\bibliographystyle{elsarticle-num}
\bibliography{}

\begin{thebibliography}{00}
\bibitem{fair}          https://www.gsi.de/en/research/fair.htm.

\bibitem{pandabook}  	The $\bar PANDA$ Coll., arXiv:0903.3905 (2009) [hep-ex].

\bibitem{pandaroot1}    D. Bertini $et~all.$, J. Phys.: Conf. Series  \textbf{119} (2008) 032011.

\bibitem{pandaroot2} S. Spataro $et~all.$, J. Phys.: Conf. Series  \textbf{396} (2012) 022048.


\bibitem{Dsbabar}   The BaBar Coll., Phys. Rev. Lett. \textbf{90} (2003) 242001.

\bibitem{dipierro1} S. Godfrey, N. Isgur,  Phys. Rev. D \textbf{32} (1985) 189. 

\bibitem{dipierro2} M. Di Pierro, E. Eichten,  Phys. Rev. D \textbf{64} (2001) 114004.

\bibitem{charm2013}  A. Palano, arXiV:1311.7364 (2013) [hep-ex].

\bibitem{Ds2460} The CLEO Coll., Phys. Rev. D \textbf{68} (2003) 032002.

\bibitem{widthgg}  The BaBar Coll., Phys. Rev. D \textbf{74} (2006) 032007.

\bibitem{purecs} S. Godfrey, Phys. Lett. B \textbf{568} (2003) 254.
\bibitem{tetraquarks} H. Y. Cheng, W. S. Hou, Phys. Lett. B \textbf{566} (2003) 193.
\bibitem{molecules} M. Cleven $et~all.$, arXiV:1405.2242 (2014) [hep-ph].

\bibitem{christoph2}  The formula is obtained in a private scientific communication with Dr. Christoph Hanhart.
\bibitem{Ds2535belle} The Belle Coll., Phys. Rev. D \textbf{77} (2008) 032001.

\bibitem{Ds2535babar} The BaBar Coll., Phys. Rev. D \textbf{83} (2011) 072003.

\bibitem{evtgen} D. J. Lange, Nucl. Instrum. Meth. A462 (2001) 152-155.

\bibitem{geant3} B. Brun, CERN-DD-EE-84-1 (1988).

\bibitem{lucao}  L. Cao and J. Ritman, J. Phys.: Conf. Series  \textbf{503} (2014) 012024.

\bibitem{belleX}   The Belle Coll., Phys. Rev. Lett. \textbf{91}  (2003) 262001.

\bibitem{yellow}   N. Brambilla $et~all.$, arXiV:0412158 (2005) [hep-ph];
N. Brambilla \textit{ et al.}, Eur. Phys. J. C \textbf{71}, 1534 (2011) arXiV:1010.5827 [hep-ph]; N. Brambilla \textit{ et al.}, arXiv:1404.3723 (2014) [hep-ph]. 

\bibitem{lhcbX}    The LHCb Coll., Phys. Rev. Lett. \textbf{110} (2013) 222001.

\bibitem{dpm-panda} A. Cappella $et~all.$, Phys. Rept. \textbf{236} (1994) 225-329.

\bibitem{martin}    M. Galuska $et~all$, PoS (Bormio 2013) 023. 

\bibitem{Xtogether} S. Lange $et~all.$, arXiV:1311.7597 (2013) [hep-ex].

\bibitem{theory3f4} T. Barnes $et~all.$, Phys. Rev. D \textbf{72} (2005) 054026

\end{thebibliography}



\end{document}